\documentclass[aps,prl,twocolumn,floatfix,superscriptaddress]{revtex4}

\usepackage{units}

\usepackage{graphicx}
\graphicspath{{figures/}}

\usepackage{xspace}
\usepackage{amsmath}

\newcommand{\micron}{\ensuremath{\unit{\mu m}}\xspace}
\renewcommand{\vec}[1]{{\boldsymbol #1}}
\newcommand{\uvec}[1]{\hat{\boldsymbol{#1}}}
\newcommand{\abs}[1]{\left\vert #1 \right\vert}

\newcommand{\real}[1]{\Re\left\{ #1 \right\}}

\begin{document}

\title{Optical forces arising from phase gradients}

\author{Yohai Roichman}

\affiliation{Department of Physics and Center for Soft Matter Research,
New York University, New York, NY 10003}

\author{Bo Sun}
\affiliation{Department of Chemistry and Center for Soft Matter
  Research, New York University, New York, NY 10003}

\author{Yael Roichman}

\author{Jesse Amato-Grill}

\author{David G. Grier}

\affiliation{Department of Physics and Center for Soft Matter Research,
New York University, New York, NY 10003}

\begin{abstract}
We demonstrate both theoretically and experimentally that
gradients in the phase of a light field exert forces
on illuminated objects, including forces transverse to the
direction of propagation.
This effect generalizes the notion of the photon
orbital angular momentum carried by helical beams of light.
We further demonstrate that these forces 
generally violate conservation of energy, and briefly discuss some
ramifications of their non-conservativity.
\end{abstract}

\pacs{radiation pressure | geometric phase | computer-generated holography |
  light-matter interaction | optical trapping}

\maketitle

Light's ability to exert forces 
has been recognized since
Kepler's \emph{De Cometis} of 1619 described the deflection of comet
tails by the sun's rays.
Maxwell demonstrated that the momentum flux in a beam of light 
is proportional to the intensity and can be
transferred to illuminated objects, resulting in
radiation pressure that pushes objects
along the direction of propagation.
This axial force has been distinguished from the transverse
torque exerted by helical beams of light carrying
orbital angular momentum (OAM) \cite{allen92}.
We demonstrate theoretically and 
confirm experimentally that OAM-induced torque is
a special case of a general class of forces
arising from phase gradients in beams of light.
We also demonstrate that phase-gradient forces
are generically non-conservative, and combine them with
the conservative forces exerted by intensity gradients to
create novel optical traps with structured force
profiles.

Our experimental demonstrations of phase-gradient forces
make use of extended optical traps created through
shape-phase holography \cite{roichman06,roichman06c,roichman07b}
in an optimized \cite{polin05} holographic optical trapping
\cite{dufresne98,grier03} system.
Holographically sculpted intensity gradients enable these generalized
optical tweezers \cite{ashkin86} to
confine micrometer-scale colloidal particles to
one-dimensional curves embedded in three dimensions.
Independent control over the intensity and phase 
profiles along the curve then provide an ideal model system
for characterizing the forces generated
by phase gradients in beams of light.

\section{Optical forces due to phase gradients}

The vector potential describing a beam of light of frequency $\omega$
and polarization $\uvec{\varepsilon}$
may be written in the form
\begin{equation}
  \label{eq:vectorpotential}
  \vec{A}(\vec{r},t) = A(\vec{r}) \, 
  \exp\left( i \, \omega t \right) \, \uvec{\varepsilon}.
\end{equation}
Assuming uniform polarization (and therefore a form of the
paraxial approximation), the spatial dependence,
\begin{equation}
  \label{eq:profile}
  A(\vec{r}) = u(\vec{r}) \, \exp\left( i \Phi(\vec{r})\right),
\end{equation}
is characterized by a non-negative real-valued amplitude,
$u(\vec{r})$, and a real-valued phase
$\Phi(\vec{r})$.
For a plane wave propagating
in the $\uvec{z}$ direction,
$\Phi(\vec{r}) = - kz$, 
where $k = n_m \, \omega/c$ is the light's wavenumber,
$c$ is the speed of light in vacuum, 
and $n_m$ is the refractive index of the medium.
Imposing a transverse phase profile $\varphi(\vec{r})$ on the wavefronts
of such a beam yields the more general form
\begin{equation}
  \label{eq:phase}
  \Phi(\vec{r}) = - k_z(\vec{r}) z + \varphi(\vec{r}),
\end{equation}
where $\uvec{z} \cdot \nabla \varphi = 0$ and where the 
the direction of the wavevector,
\begin{equation}
  \label{eq:wavevector}
  \vec{k}(\vec{r}) = k_z(\vec{r}) \, \uvec{z} + \nabla \varphi,
\end{equation}
now varies with position.
% in the beam
%under the constraint 
%$k^2 = \abs{\vec{k}}^2 = k_z^2 + \abs{\nabla \varphi}^2$. 
%This is appropriate in the paraxial 
%limit $k \gg \abs{\nabla \varphi}$.
The associated electric and magnetic fields
are given in the Lorenz gauge by
\begin{align}
  \vec{E}(\vec{r},t) & = -\frac{\partial \vec{A}(\vec{r},t)}{\partial t}
  \quad \text{and}\\
  \vec{H}(\vec{r},t) & = \frac{1}{\mu} \, \nabla \times \vec{A}(\vec{r},t),
\end{align}
where $\mu$ is the magnetic permeability of the medium, which we
assume to be linear and isotropic.
Following the commonly accepted Abraham formulation \cite{loudon04},
the momentum flux carried by the beam is
\begin{align}
  \label{eq:poynting}
  \vec{g}(\vec{r}) & = \frac{1}{c} \, \real{\vec{E}^\ast \times \vec{H}} \\
  & = \frac{k}{\mu n_m} \, I(\vec{r}) \, \nabla \Phi,
  \label{eq:momentumflux}
\end{align}
where $I(\vec{r}) = u^2(\vec{r})$ is the light's intensity, and
where we have employed the gauge condition $\nabla \cdot \vec{A} = 0$.
This result is somewhat controversial \cite{leonhardt06} despite a century of study,
with the alternative Minkowski formulation
differing from Eq.~(\ref{eq:momentumflux})
by a factor of $n_m^2$.
Both formulations are believed to yield consistent results for the
force on an illuminated object, but differ in their predictions
for the recoil of the medium \cite{loudon04}.

Applying Eq.~(\ref{eq:phase}), $\vec{g}(\vec{r})$ separates naturally
into an axial component
$\vec{g}_z(\vec{r}) = [k I(\vec{r}) \, k_z/(\mu n_m)] \, \uvec{z}$
and an additional component
\begin{equation}
  \label{eq:transverse}
  \vec{g}_\perp(\vec{r}) = \frac{k}{\mu n_m} \, I(\vec{r}) \,
  \nabla \varphi
\end{equation}
transverse to the optical axis.
For small objects,
the transverse force arising from $\vec{g}_\perp(\vec{r})$
may be approximated as
\begin{equation}
  \label{eq:phaseforce}
  \vec{f}_\perp(\vec{r}) = \sigma(a) \, \vec{g}_\perp(\vec{r}),
\end{equation}
where $\sigma(a)$ is the cross-section 
describing the light-matter interaction.
For a dielectric sphere of radius $a$, \cite{ashkin86,harada96c},
\begin{equation}
  \label{eq:crosssection}
  \sigma(a) = \frac{8 \pi}{3} \, k^4 a^6 \,
  \left(\frac{m^2 - 1}{m^2 + 2}\right),
\end{equation}
where $m = n_p/n_m$ is the ratio of the particle's refractive
index, $n_p$, to the medium's, $n_m$.

More than a decade ago, Allen, \emph{et al.} 
\cite{allen92} pointed out that
the helical phase profile,
$\varphi_\ell(\vec{r}) = \ell \theta,$
imbues a beam of light with an orbital angular momentum flux,
$\vec{r} \times \vec{g}_\perp$,
amounting to $\ell \hbar$ per photon.
Here, $\theta$ is the azimuthal angle around the optical axis, and
$\ell$ is an integer describing the wavefronts' helical pitch.
This orbital angular momentum is distinct from the
photons' intrinsic spin angular momentum 
\cite{allen92,simpson97,allen99,oneil00,oneil02,curtis03}.
Through it, even linearly polarized light can exert a torque
around the optical axis \cite{he95a,friese96,simpson96}.
Equations~(\ref{eq:transverse}) and~(\ref{eq:phaseforce})
reveal the torque exerted by helical wavefronts
to be a particular manifestation
of the more general class of transverse forces
arising when phase
gradients redirect radiation pressure.

Intensity gradients also exert forces on illuminated objects
\cite{ashkin86}.
In this case, the dipole moment induced in the object
responds to gradients in the field, yielding a force
proportional to the gradient of the intensity.
For a small sphere, this has the form
\cite{ashkin86,harada96c},
\begin{equation}
  \label{eq:gradientforce}
  \vec{F}_\nabla(\vec{r}) = n_m \, \frac{k^2 a^3}{2} \, 
  \left(\frac{m^2 - 1}{m^2 + 2}\right) \, \nabla I.
\end{equation}
Unlike the phase-gradient force, $\vec{F}_\nabla$ can be
directed up the optical axis, and so can be used to
form stable three-dimensional optical traps, despite
axial radiation pressure.
This is the basis for optical tweezers and related single-beam
optical traps \cite{ashkin86}.

Forces due to intensity gradients are manifestly conservative
\cite{ashkin92},
arising as they do from the gradient of an analytic function.
Phase-gradient forces, by contrast, generally are not conservative.
This is evident because
\begin{equation}
  \label{eq:nonconservative}
  \nabla \times \vec{g} = \frac{k}{\mu n_m} \, (\nabla I) \times
  (\nabla \Phi) \neq 0,
\end{equation}
in general.
This noteworthy property of scattering forces
was pointed out by Ashkin \cite{ashkin92},
but appears not to have been emphasized subsequently
in the optical trapping literature.
Effects arising from non-conservative optical
forces may not have drawn much attention
because the experimentally accessible transverse
forces can appear to be conservative in the
symmetric beams that usually are considered.
The phase-gradient force in a helical beam, by contrast,
accelerates objects around the optical axis 
and so obviously does not conserve energy.

\section{Creating optical traps with phase gradients}

\begin{figure}[!t]
  \centering
  \includegraphics[width=\columnwidth]{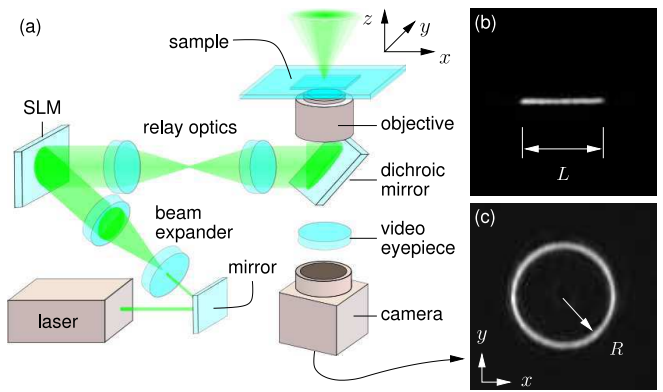}
  \caption{Schematic representation of the optimized holographic
    optical trapping system using shape phase holography to project
    extended optical traps.}
  \label{fig:schematic}
\end{figure}

We demonstrate and characterize phase-gradient forces
using the optimized holographic optical trapping technique
\cite{polin05,dufresne98,roichman05a} and
shape-phase holography \cite{roichman06,roichman07b}
to create extended optical traps with
specified profiles for both their intensity and phase.
Our apparatus, shown schematically in Fig.\ \ref{fig:schematic}(a),
uses a phase-only spatial light modulator (Hamamatsu X7690-16)
to imprint computer-generated holograms on the wavefronts
of a slightly converging \cite{polin05} beam of light provided
by a solid-state laser (Coherent Verdi 5W) operating at
a vacuum wavelength of 532~\unit{nm}.
The modified beam is relayed to an objective lens
(Nikon Plan Apo, $100\times$ oil immersion, NA 1.4)
that focuses it into the intended three-dimensional optical trapping pattern.
This objective lens is mounted in an inverted optical microscope
(Nikon TE-2000U), and the trapping beam is directed into
the objective's input pupil with a wavelength selective polarizing
beam splitter.
Light at other wavelengths passes through the beam splitter
to form images on a CCD camera (NEC TI-324AII).

The holograms used for this study are designed 
to bring the light to a
focus along one-dimensional curves, $C$, embedded in the
three-dimensional focal volume of the objective lens.
The hologram also encodes a designated 
intensity profile $I(s)$ and phase profile $\varphi(s)$
along the arclength $s$ of $C$.
This is accomplished by numerically back-projecting 
\cite{goodman05} the intended field along $C$ back to the plane of 
the SLM to obtain the ideal complex-valued hologram,
$\Psi(\vec{r}) = b(\vec{r}) \, \exp(i p(\vec{r}))$.
The real-valued amplitude $b(\vec{r})$ is defined to be non-negative
by absorbing sign changes into the definition of the
phase, $p(\vec{r})$.
This facilitates encoding the complex hologram on a
phase-only SLM \cite{roichman06,roichman06c,roichman07b}.
The shape-phase algorithm
assigns $p(\vec{r})$ to the
phase pixel at $\vec{r}$ with a probability proportional
to $b(\vec{r})$.  
An alternate phase pattern imprinted on unassigned pixels
is used to divert
excess light away from the intended trapping pattern.
How the SLM plane is divided into assigned and unassigned
domains can be tuned to minimize artifacts, improve
diffraction efficiency, and optimize accuracy 
\cite{roichman06,roichman06c,roichman07b}.

The images in Figs.~\ref{fig:schematic}(b) and (c)
show a focused line trap \cite{roichman06}
and ring trap, respectively, each designed to have uniform 
intensity and uniform phase gradients.
These images were obtained by placing a mirror in the
microscope's focal plane and capturing an image of the
reflected light.
Because these extended traps come to a diffraction-limited
focus, their axial intensity gradients are steep enough
to trap particles stably in three-dimensions against
radiation pressure \cite{ashkin86}.

\section{Phase gradient forces in line traps}

\subsection{Linear phase gradients}
\begin{figure}[!t]
  \centering
  \includegraphics[width=\columnwidth]{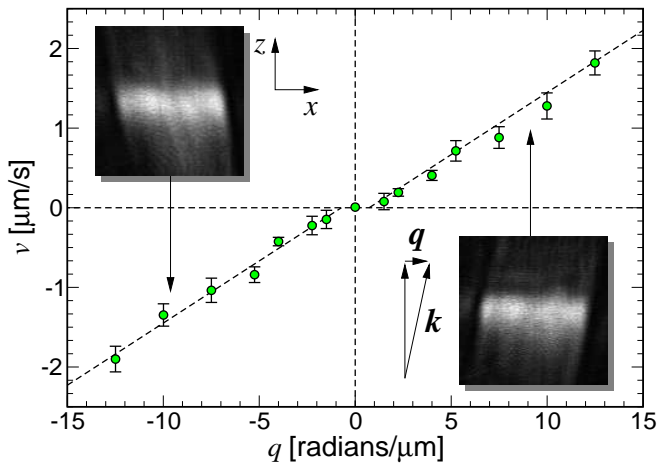}  
  \caption{Linear dependence of a particle's mean velocity $v$ along a uniformly bright line tweezer of length $L$ on phase gradient $q$.}
  \label{fig:vdv}
\end{figure}

To demonstrate the phase gradient force predicted by 
Eq.~(\ref{eq:transverse}), we used digital video
microscopy \cite{crocker96} to track micrometer-scale
colloidal spheres moving along uniformly bright
line traps such as that in Fig.~\ref{fig:schematic}(b).
The trap-forming hologram was adaptively optimized
to minimize intensity variations
and to incorporate linear
phase gradients, $\nabla \varphi = q \uvec{x}$,
ranging from $q = -12\,\unit{radians/\micron}$ to $q =
12\,\unit{radians/\micron}$.
The effect of imposing this transverse phase gradient can be seen
in axial sections \cite{roichman06c} through the projected
three-dimensional intensity distributions, shown
inset into Fig.~\ref{fig:vdv}.
The diffraction-limited focal line remains
in the $xy$ plane despite the imposed phase gradient.
The beam's direction of propagation, however, deviates from the $\uvec{z}$
axis by the angle $\sin^{-1}(q/k)$.
This tilt directs a component of the beam's wave vector
along the line, thereby creating a longitudinal component
of the radiation pressure.
The measured angle of inclination
and its uniformity along $x$
confirm both the phase gradient's magnitude and also its
uniformity.

Line traps with tunable linear phase gradients 
were projected into aqueous dispersions of
colloidal silica spheres $2a = 1.53~\micron$ in diameter
sealed into the
40~\micron thick gap
between a glass microscope slide and a \#1 glass coverslip.
The line traps were focused near the samples' midplane
to minimize both interference with reflections from the glass-water
interface
and also
hydrodynamic coupling to the walls.
In that case, the spheres' viscous drag coefficient is
reasonably approximated \cite{dufresne01}
by the Stokes value $\gamma = 6 \pi \eta a$,
where $\eta = 1~\unit{cP}$ is the viscosity of water.
Equations~(\ref{eq:transverse}) and (\ref{eq:phaseforce}) then
suggest that a single sphere's speed, $v$, along the line
should be proportional to $q$.

To test this prediction, we
measured the time required for a single sphere 
to travel the line's length, $L$, as the sign of $q$ was
flipped 20 times for each value of $\abs{q}$.
An estimated 100~\unit{mW} of light formed the
trap, after accounting for losses in the optical train.
The observed off-line excursions of roughly 200~\unit{nm}
(root-mean-squared) suggest an axial and lateral trap
stiffness comparable to that of a point-like optical tweezer
powered by 1~\unit{mW}.
Under these conditions, the trapped sphere
traveled the length of the line trap
in roughly 2~\unit{s} when subjected to the largest phase
gradients.
Results obtained by systematically varying $q$ are
plotted in Fig.~\ref{fig:vdv}, and clearly show
the anticipated linear dependence, except very near $q = 0$.
The smallest values of $\abs{q}$ created too weak a force
to overcome
localized traps formed by uncorrected intensity variations.
%Similar results are obtained with the polarization vector
%along or transverse to the line.

\subsection{Phase-gradient traps}
\begin{figure}
  \centering
  \includegraphics[width=\columnwidth]{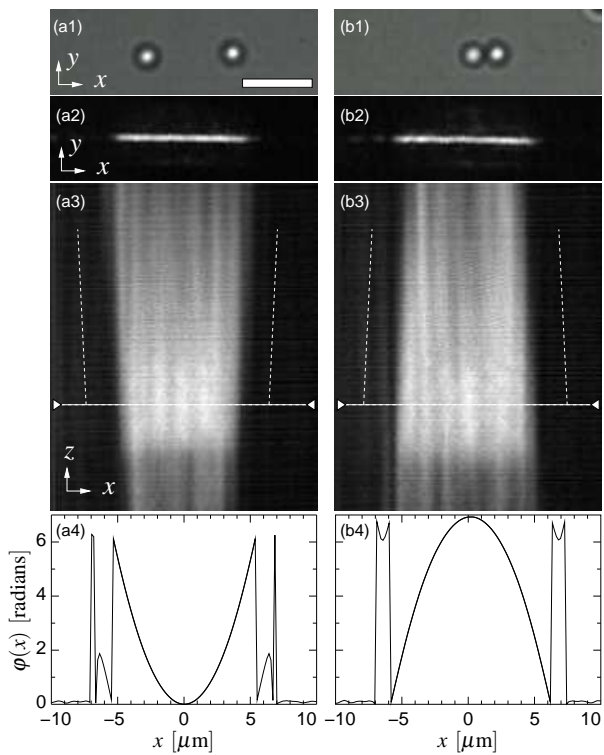}
  \caption{Phase-gradient potential energy barrier (a)
    and well (b) created along a uniformly bright line trap.
    (1) Two 1.5~\micron diameter silica spheres trapped on the line.
    (2) The uniform in-plane intensity of the focused line.
    (3) Axial section through
    the measured intensity profile, show the divergence (a3) and
    convergence (b3) of the beams created by the imposed
    phase profile.
    (4) The designed phase gradient, featuring the designed
    parabolic profile and off-line phase variations designed to
    minimize intensity variations.  
}
  \label{fig:phasewell}
\end{figure}

More complicated phase gradients give rise to
more interesting physical effects.
The particles shown in Fig.~\ref{fig:phasewell} also 
are trapped along a uniformly bright line trap of 
length $L = 10~\micron$.
This line, however, has a parabolic phase profile along its length,
$\varphi(x) = \pm (qx)^2$ that is predicted to create either
force objects toward the center of the line, or out to the ends,
depending on the choice of sign.
The data in Fig.~\ref{fig:phasewell} also confirm this prediction.
Axial sections through the projected intensity pattern
show that the phase-gradient barrier results from light diverging along
the line's length, 
while the well results from the projection of converging rays
onto the trap's length.
So long as the particles remain rigidly confined to the
uniformly bright focal line, the phase-gradient force approximates
a conservative one-dimensional potential energy landscape.

\section{Phase gradient forces in ring traps}
\begin{figure}
  \centering
  \includegraphics[width=\columnwidth]{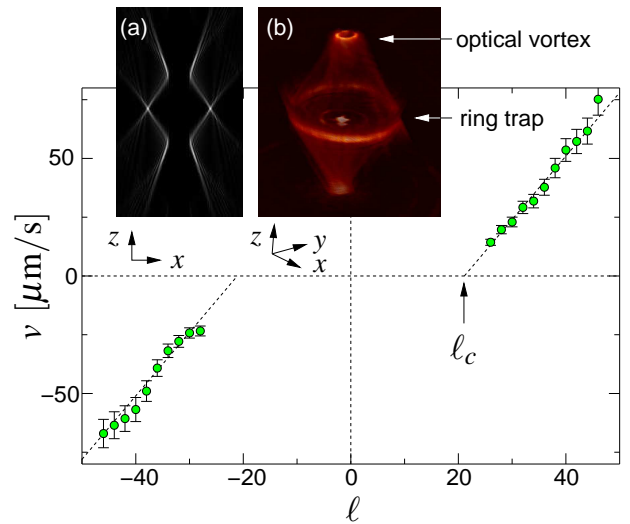}
  \caption{Colloidal transport driven by azimuthal phase
    gradients in holographic ring traps.  (a) Computed axial
    section through a holographic ring trap of radius $R = 20~\micron$
    and helicity $\ell = 30$.  (b) Volumetric representation of
    the measured three-dimensional intensity field in a holographic
    ring trap of radius $R = 20~\micron$ and $\ell = 10$.
    Data points show the peak speed $v$ of a single colloidal silica
    sphere circulating around the ring in (b) as a function of
    topological charge $\ell$.}
  \label{fig:collapse}
\end{figure}

Like holographic line traps, holographic ring traps, such as
the example in Fig.~\ref{fig:schematic}(c), can be endowed with
arbitrary phase profiles, including the uniform azimuthal phase
gradient, $\varphi(\vec{r}) = \ell \theta$, that defines a helical mode.
A helical profile, by itself, causes a beam to focus into a
ring of light, forming a torque-exerting optical trap known
as an optical vortex \cite{simpson96,he95,gahagan96}.
Whereas the radius of an optical vortex is proportional to its
helicity \cite{curtis03,sundbeck05}, holographic ring traps
can be projected with any desired radius, independent of $\ell$.
This facilitates systematic studies of colloidal transport
under varying phase gradients.
Also unlike optical vortices, holographic ring traps come to
a diffraction-limited focus, independent of $\ell$, and therefore
can trap micrometer-scale objects stably in three dimensions.

This property can be seen in the computed
axial section through a holographic
ring trap shown in Fig.~\ref{fig:collapse}(a).
The section through the trap itself appears as two
bright focal spots at the images mid-line.
The ring's axial intensity gradients are comparable
to those of a conventional optical tweezer, and
so suffice to trap objects stably in three dimensions.
This ring trap features a helicity of $\ell = 30$,
and so is expected to exert a torque around the optical
axis proportional to $\ell$.
The helical wavefront topology also
suppresses the beam's axial intensity
through destructive interference.  Consequently, none of the
light projected by the objective lens appears along the beam's axis.
Rather, it is diverted to a radius, $R_\ell$ from the axis
\cite{curtis03,sundbeck05}.  Provided that the
ring trap's radius, $R$, exceeds $R_\ell$,
the converging helical beam focuses not only to the intended
ring trap in the focal plane, but also to two 
conventional optical vortices above and below the ring
trap.
Sections through these optical vortices appear as bright
regions above and below the mid-line in Fig.~\ref{fig:collapse}(a).
Neither features the strong
axial intensity gradients of the holographic ring.
Consequently, neither can trap objects in three dimensions
against radiation pressure.
This structure also is evident in the three-dimensional 
reconstruction \cite{roichman06c}
of an experimentally projected holographic ring trap 
in Fig.~\ref{fig:collapse}(b).

We measured the free-running speed of a single colloidal silica sphere
of radius $a = 0.76~\micron$ as it
circulated around holographic ring traps of radius $R = 2.6~\micron$
projected into the midplane of a 40~\micron thick sample.
The trapped particle was subjected to azimuthal phase
gradients ranging from $\ell = -50$ to $\ell = 50$, and
its peak speed was measured \cite{crocker96}
to within 10\% for each value of the helicity.
The results are plotted in Fig.~\ref{fig:collapse}.

Like optical vortices, holographic ring trap are subject
to $\ell$-fold and $2\ell$-fold azimuthal intensity
variations due to phase scaling errors \cite{lee05c}.
These create localized traps that compete
with the intended phase-gradient force
\cite{curtis03,lee06a}.
Consequently, the particle remains motionless for helicities
below a threshold, $\ell = \ell_c$.
Similar pinning due to intensity gradients appears in
Fig.~\ref{fig:vdv}.
For $\ell > \ell_c$, however, the particle's peak speed
increases linearly with $\ell$, consistent with
the predictions of Eqs.~(\ref{eq:transverse}) and
(\ref{eq:phaseforce}).
Intermittent circulation near $\ell = \ell_c$ gives rise
to large velocity fluctuations characterized by giant
enhancement of the particle's effective diffusion coefficient
\cite{lee06a}.
Disorder in the effective force landscape also gives rise to
interesting collective dynamics for multiple particles
trapped on the ring, including transitions among periodic,
chaotic and weakly chaotic steady states \cite{roichman07}.
Phase-gradient forces in holographic ring traps therefore
provide useful model systems for studying fundamental 
problems in nonequilibrium statistical mechanics.
They also promise practical applications as the basis for
microscopic pumps \cite{ladavac04a}, mixers \cite{curtis02},
and optomechanical micromachines \cite{ladavac05}.

\section{Conclusions}

The foregoing sections demonstrate that phase gradients in beams
of light give rise to transverse forces arising from the
redirection of radiation pressure.
The existence of orbital angular momentum in beams of light and its
transfer to illuminated objects can be seen to be manifestations
of this general effect.

Although phase-gradient forces
may appear to be conservative on
restricted manifolds, they are
not conservative in general.
Non-conservative optical forces may give rise to interesting
effects in illuminated particles' dynamics, including
departures from Boltzmann statistics for systems nominally
in equilibrium.
Such effects may substantially affect measurements
of microscopic interactions and dynamics based on
the statistics of optically trapped particles.
They appear not to have been addressed in the large
and rapidly growing literature in this area.

We have demonstrated that
phase-gradient forces can be used to craft one-dimensional
force profiles in uniformly bright holographic line traps.
These can be used to probe the interactions between multiple
particles trapped on a line,
including both intrinsic colloidal interactions
and also light-induced interactions.
This application for phase-gradient
forces in holographic line traps will be discussed elsewhere.

The phase-gradient force in holographic line
traps is subject to the Abraham-Minkowski controversy.
Extended line traps created with shape-phase holography
therefore may
provide useful model systems for probing how a medium
influences the momentum flux of light.

Finally, it is worth mentioning that the polarization
generally will vary with position in beams of light with
spatially varying phase.  Although 
optically isotropic materials are not
influenced by polarization gradients,
anisotropic materials can be.
Polarization gradients therefore should provide
additional independent avenues for exerting controlled
forces on some microscopic systems.

\begin{acknowledgments}
This work was supported by the National Science Foundation through
Grant Number DMR-0606415.  We are grateful to Marco Polin for
enlightening conversations and to Andy Hollingsworth for 
help in synthesizing and characterizing
monodisperse colloidal silica spheres.
\end{acknowledgments}

%\bibliographystyle{pnas}
%\bibliography{abbreviations,grier,dgg,tweezer}

\end{document}